\documentclass[aps,pra,twocolumn,showpacs,superscriptaddress]{revtex4}
\usepackage{amsmath}
\usepackage{mathrsfs}
\usepackage{amssymb}
\usepackage{amsfonts}
\usepackage{graphicx}

\begin{document}

\title{Reduced fidelity approach for quantum phase transitions in spin-1/2
dimerized Heisenberg chains}
\author{Heng-Na Xiong}
\affiliation{Zhejiang Institute of Modern Physics, Department of Physics, Zhejiang
University, HangZhou 310027, People's Republic of China.}
\author{Jian Ma}
\affiliation{Zhejiang Institute of Modern Physics, Department of Physics, Zhejiang
University, HangZhou 310027, People's Republic of China.}
\author{Zhe Sun}
\affiliation{Department of Physics, HangZhou Normal University, HangZhou 310036, People's
Republic of China.}
\author{Xiaoguang Wang}
\email{xgwang@zimp.zju.edu.cn}
\affiliation{Zhejiang Institute of Modern Physics, Department of Physics, Zhejiang
University, HangZhou 310027, People's Republic of China.}
\date{\today}

\begin{abstract}
We use reduced fidelity approach to characterize quantum phase transitions
in the one-dimensional spin-1/2 dimerized Heisenberg chain in the
antiferromagnetic case. The reduced fidelity susceptibilities between two
nearest-neighboring spin pairs are considered. We find that they are
directly related to the square of the second derivative of the ground-state
energy. This enables us to conclude that the former might be a more
effective indicator of the second-order quantum phase transitions than the
latter. Two further exemplifications are given to confirm the conclusion is
available for a broad class of systems with SU(2) and translation
symmetries. Moreover, a general connection between reduced fidelity
susceptibility and quantum phase transitions is illustrated.
\end{abstract}

\maketitle

\section{Introduction}

Quantum phase transitions (QPTs) is an essential phenomenon in quantum
many-body correlated system. It is induced by the ground-state (GS)
transition driven by external parameters at zero temperature. How to
characterize QPTs has attracted widespread attention. Conventionally, QPTs
are described in terms of order parameter and symmetry breaking within the
Landau-Ginzburg paradigm \cite{Book-QPT}. One object in these traditional
ways is that there is no general method to find the order parameter for a
common system. To overcome this problem, a concept called \emph{fidelity}
\cite{fidelity-quan, fidelity-zanardi} (see Eqs.~(\ref%
{eq-definition-pure-fidelity}) and (\ref{eq-definition-mixed-fidelity})) is
borrowed from the field of quantum-information theory, since it well
describes the overlap between two states in different phases with different
structural properties, i.e., order parameters. Thus it dose not need a
priori knowledge of the order parameter in detecting QPTs. It is a purely
Hilbert-space geometrical quantity. On the other hand, \emph{fidelity
susceptibility} \cite{fidelity-susceptbility-Gu,fidelity-susceptbility-paolo}
(see Eq.~(\ref{eq-definition-fidelity-susceptibility})) is found more
convenient than fidelity itself for its independence of the slightly changed
external parameters. Hitherto, these two connected concepts have succeeded
in identifying the QPTs of many systems, such as XY spin chains and Dicke
model \cite{fidelity-zanardi}, XXZ chain \cite{fidelity-XXZ-model}, Hubbard
model \cite{fidelity-hubbard-model-1,fidelity-hubbard-model-2}, frustrated
Heisenberg chain \cite{fidelity-frustrated-model}, Kitaev honeycomb model
\cite{fidelity-Kitav-model}, extended Harper model \cite%
{fidelity-Harper-model}. The intrinsic relation between the GS fidelity (or
fidelity susceptibility) and the characterization of a quantum phase
transition has been unveiled in Ref. \cite{fidelity-energy-chen}. It was
shown that the singularity and scaling behavior of the GS fidelity (or
fidelity susceptibility) are directly related to its corresponding
derivative of GS energy, which characterizes the QPTs conventionally.
Moreover, the fidelity susceptibility is associated with dynamic structure
factor for QPTs and with specific heat and magnetic susceptibility for
thermal phase transitions \cite{fidelity-dynamicstructurefactor}.

The above works are all concerned with the global GS fidelity. Then there is
a natural question that whether the fidelity of the subsystem, i.e., the
\emph{reduced fidelity} (or named \emph{partial-state fidelity}) could
reflect the QPTs. Recently, some works have been devoted to this subject.
Zhou \textit{et al}. \cite{averaged-fidelity-zhou} found that it succeeds in
capturing nontrivial information along renormalization group flows and in
detecting the QPTs in XY model \cite{averaged-fidelity-zhou-XYmodel}.
Paunkovi\'{c} \textit{et al}. \cite{partial-fidelity} showed that it enables
them to identify the on-site magnetization as the order parameter for the
phase transition in the conventional BCS superconductor with an inserted
magnetic impurity system. Kwok \textit{et al}. \cite%
{partial-fidelity-LMG-model} tested its effectiveness in charactering the
QPTs of the isotropic Lipkin- Meshkov-Glick model and the antiferromagnetic
one-dimensional Heisenberg model. Meanwhile, we derived a general expression
for the two-site reduced fidelity susceptibility (RFS). It has been applied
to the study of the Lipkin-Meshkov-Glick model \cite{reduced-fidelity-LMG-Ma}
and transverse field Ising model \cite{reduced-fidelity-TIsing-Ma}. We found
that the RFS shows similar scaling behavior to the global fidelity
susceptibility. All the above works illustrate that the reduced fidelity
approach is also an effective tool in identifying QPTs. However, a general
relation between RFS and QPTs is not established.

In this work, we apply the reduced fidelity approach to the one-dimensional
(1D) spin-1/2 Heisenberg chain with dimerization. Thanks to the SU(2) and
translation symmetries, we derive general expressions of the two-site RFSs
(the RFSs we mentioned below are all for two-site.), which are connected
closely to the square of the second derivative of the GS energy. This result
indicates that the RFS is an more effective tool to identify the
second-order QPTs than the second derivative of the GS energy. To further
testify our conclusion, we exemplify the mixed-spin dimerized Heisenberg
chain and the spin-1 bilinear-biquadratic model as well. These two models
are both of SU(2) and translation symmetries too. Furthermore, it is
illustrated that, in general, the origin for RFS to signal QPTs may root in
the relation between the reduced density matrix (RDM) and the derivatives of
GS energy.

This paper is organized as follows. In Sec.~II, we derive a general
expression of RFS for two Hermitian and semi-positive definite density
matrices, which are commute with each other, and give a direct connection
between RFSs and QPTs in the dimerized model. In Sec.~III, the critical
behavior of the system is studied for both finite-size and infinite-size
situations. In Sec.~IV, two further models are enumerated and a possible
origin between the relation of general reduced fidelity and QPTs is
illustrated. Finally, a summary is presented in Sec.~V.

\section{Reduced fidelity susceptibility and its connection to quantum phase
transitions}

The dimerized Heisenberg chain is a fundamental spin-correlated model. It is
of special interest both in theory and experiment, since it gives a
reasonably accurate description of many quasi-1D antiferromagnets which have
two important but structurally inequivalent superexchange paths that are
spatially linked, such as the materials of Cu$\left( \text{NO}_{3}\right)
_{2}\cdot $2.5H$_{2}$O$,$ $\left( \text{VO}\right) _{2}$P$_{2}$O$_{7}$ and
various aromatic free-radical compounds \cite{DM-materials}. Therefore, many
efforts have been devoted to study its quantum critical behavior of the
dimerized Heisenberg model using various methods, e.g., continuous unitary
transformations \cite{DM-QPT-CUT}, density matrix renormalization group \cite%
{DM-QPT-DMRG}, concurrence \cite{DM-QPT-concurrence} and block entanglement
\cite{DM-QPT-entanglement}. Here we employ the reduced fidelity approach to
study the QPTs of this model.

The Hamiltonian for antiferromagnetic Heisenberg chain (AHC) with
dimerization reads
\begin{equation}
H_{D}=\sum_{i=1}^{N/2}\left( \mathbf{S}_{2i-1}\cdot \mathbf{S}_{2i}+\alpha
\mathbf{S}_{2i}\cdot \mathbf{S}_{2i+1}\right) ,  \label{Hamiltonian}
\end{equation}%
where ${\mathbf{S}}_{i}$ denotes the $i$-th spin-1/2 operator, and $\alpha
>0 $ is the ratio between the two kinds nearest-neighboring (NN) couplings.
The total number of spins $N$ is required to be even and the periodic
boundary condition $\mathbf{S}_{1}=\mathbf{S}_{N+1}$ is assumed.

\subsection{Reduced density matrix}

To study the RFS, we need to know the RDM between two spins, and through the
whole discussion we restrict to the case of two NN spin pairs. The
Hamiltonian has the SU(2) symmetry, i.e., $\left[ H,\sum_{i=1}^{N}S_{i\gamma
}\right] =0$ $\left( \gamma =x,y,z\right) ,$ which guarantees the RDM
between two NN spins is of the form~\cite{RDM-from}
\begin{equation}
\rho _{ij}=\text{diag}\left( \varrho _{1},~\varrho _{2}\right) ,
\label{eq-dimer-RDM-1}
\end{equation}%
with%
\begin{equation}
\varrho _{1}=%
\begin{pmatrix}
u^{+} & 0 \\
0 & u^{+}%
\end{pmatrix}%
,\text{ }\varrho _{2}=%
\begin{pmatrix}
u^{-} & w \\
w & u^{-}%
\end{pmatrix}%
,  \label{eq-dimer-RDM-2}
\end{equation}%
in the basis $\left\{ |00\rangle ,|11\rangle ,|01\rangle ,|10\rangle
\right\} $, where $\sigma _{z}|0\rangle =-|0\rangle $ and $\sigma
_{z}|1\rangle =|1\rangle $. The matrix elements are given by~\cite{RDM-from}
\begin{eqnarray}
u^{\pm } &=&\frac{1}{4}\left( 1\pm \langle \sigma _{iz}\sigma _{jz}\rangle
\right) ,  \notag \\
w &=&\frac{1}{2}\langle \sigma _{iz}\sigma _{jz}\rangle .
\label{eq-elements-RDM}
\end{eqnarray}%
This implies the RDM $\rho _{ij}$ is only related to the spin correlator $%
\langle \sigma _{iz}\sigma _{jz}\rangle $. It is noticed that both $\varrho
_{1}$ and $\varrho _{2}$ are Hermitian, and they can be rewritten in terms
of Pauli operators as $\varrho _{1}=u\mathbf{I},\text{ \ \ \ }\varrho _{2}=w%
\mathbf{I}+z\sigma _{x},$ where $\mathbf{I}$ denotes a $2\times {2}$
identity matrix. Therefore, it is found that $\varrho _{i}\equiv \varrho
_{i}\left( \alpha \right) $ $\left( i=1,2\right) $ commutes with $\widetilde{%
\varrho }_{i}\equiv \varrho _{i}\left( \alpha +\delta \right) \,$\ with $%
\delta $ a small perturbation of the control parameter $\alpha $, i.e., $%
[\varrho _{i},~\widetilde{\varrho }_{i}]=0$. This commuting property will
great facilitate our study of RFS below.

In addition, there is an translational invariance in the Hamiltonian due to
the periodic boundary condition, which leads to the fact that any two terms
of the form $\langle \mathbf{S}_{i}\cdot \mathbf{S}_{i+1}\rangle $ equals to
each other. Applying the Feynman-Hellman theorem \cite{Feynman-Hellman
theorem}, i.e., $\partial _{\alpha }E_{n}=\langle {n}|\partial _{\alpha }H|{n%
}\rangle $ with $|n\rangle $ the non-degenerate eigenstate of Hamiltonian $H$
and $E_{n}$ the eigenenergy. To the GS, the spin correlators corresponding
to two NN spin pairs are written as
\begin{eqnarray}
\langle \sigma _{1z}\sigma _{2z}\rangle  &=&\frac{8}{3}({e_{0}}-{\alpha }{%
\partial _{\alpha }e_{0}}),\text{ }  \notag \\
\langle \sigma _{2z}\sigma _{3z}\rangle  &=&\frac{8}{3}{\partial _{\alpha
}e_{0}},  \label{spin-correlators}
\end{eqnarray}%
where $e_{0}\equiv {E_{0}/N}$ represents the GS energy (denoted by $E_{0}$)
per spin. The above equation gives a direct relation between the spin
correlators and the GS energy and its first derivative. In other words, the
elements of the RDMs are completely determined by $e_{0}$ and $\partial
_{\alpha }e_{0}$.

\subsection{Reduced fidelity susceptibility}

First, we briefly review the definitions of fidelity and fidelity
susceptibility. For two pure states $|{\Psi (\alpha )}\rangle $ and $|{\Psi
(\alpha +\delta )}\rangle $ with $\delta $ a small change of the external
parameter $\alpha $, their overlap or \emph{fidelity} is defined as \cite%
{fidelity-zanardi}
\begin{equation}
F(\alpha )=|\langle {\Psi (\alpha )}|{\Psi (\alpha +\delta )}\rangle |.
\label{eq-definition-pure-fidelity}
\end{equation}%
The extension to the mixed states is in general the \emph{Uhlmann fidelity}
\cite{Uhlmann-fidelity-Uhlmann,Uhlmann-fidelity-Jozsa}
\begin{equation}
F(\alpha )=\text{tr}\sqrt{\rho (\alpha )^{1/2}\rho (\alpha +\delta )\rho
(\alpha )^{1/2}},  \label{eq-definition-mixed-fidelity}
\end{equation}%
with $\rho (\alpha )$ and $\rho (\alpha +\delta )$ the two density matrices.
The fidelity susceptibility is defined as
\begin{equation}
\chi =\text{lim}_{\delta {\rightarrow {0}}}\frac{-2\ln {F}}{\delta ^{2}}.
\label{eq-definition-fidelity-susceptibility}
\end{equation}%
Thus the fidelity susceptibility does not depend on $\delta $.

Then, we will generally calculate the fidelity between two Hermitian and
semi-positive definite density matrices $\varrho \equiv \varrho \left(
\alpha \right) $ and $\widetilde{\varrho }\equiv \varrho \left( \alpha
+\delta \right) ,$ which are commute with each other, i.e., $\left[ \varrho
,~\widetilde{\varrho }\right] =0$, so that they can be diagonalized
simultaneously. With the definition of fidelity, we get%
\begin{equation}
F_{\varrho }=\text{tr}\sqrt{\varrho ^{1/2}\widetilde{\varrho }\varrho ^{1/2}}%
=\sum_{i}\sqrt{\lambda _{i}\widetilde{\lambda }_{i}},
\label{eq-RF-general-expression}
\end{equation}%
where $\lambda _{i}$s and $\widetilde{\lambda }_{i}$s are the eigenvalues of
$\varrho $ and $\widetilde{\varrho }$, respectively. Since zero eigenvalues
have no contribution to $F_{\varrho }$, we only need to consider the nonzero
ones. In the following, the subscript $i$ in $\sum_{i}$ only refers to the
nonzero eigenvalues of $\varrho $.

For a small change $\delta ,$ $\widetilde{\lambda }_{i}$ can be expanded as $%
\widetilde{\lambda }_{i}\equiv {\lambda (\alpha +\delta )}\simeq \lambda
_{i}+\left( \partial _{\alpha }\lambda _{i}\right) \delta +\left( \partial
_{\alpha }^{2}\lambda _{i}\right) \delta ^{2}/{2}+O\left( \delta ^{3}\right)
$. Then the fidelity for matrix $\varrho $ becomes
\begin{equation}
F_{\varrho }=\sum_{i}\left\{ \lambda _{i}+\frac{\delta }{2}\partial _{\alpha
}\lambda _{i}+\frac{\delta ^{2}}{4}\left( \partial _{\alpha }^{2}\lambda
_{i}-\frac{\left( \partial _{\alpha }\lambda _{i}\right) ^{2}}{2\lambda _{i}}%
\right) \right\} .
\end{equation}%
Here we have neglected small terms higher than second order. Since $%
\sum_{i}\lambda _{i}\equiv {1},$ we have $\sum_{i}\partial _{\alpha }\lambda
_{i}=\sum_{i}\partial _{\alpha }^{2}\lambda _{i}=0$. Thus the fidelity is
further reduced to
\begin{equation}
F_{\varrho }=1-\frac{\delta ^{2}}{2}\sum_{i}\frac{\left( \partial _{\alpha
}\lambda _{i}\right) ^{2}}{4\lambda _{i}}.
\end{equation}%
Therefore, according to the relation between fidelity and susceptibility, $%
F=1-\chi {\delta ^{2}/2}$, which is equivalent to Eq.~(\ref%
{eq-definition-fidelity-susceptibility}), the fidelity susceptibility $\chi
_{\varrho }$ corresponding to the matrix $\varrho $ is obtained as
\begin{equation}
\chi _{\varrho }=\sum_{i}\frac{\left( \partial _{\alpha }\lambda _{i}\right)
^{2}}{4\lambda _{i}}.  \label{fs}
\end{equation}%
This expression of fidelity susceptibility is valid for any commuting
density matrices, and the second power on the right-hand side of the
equation will lead to an interesting relation between the RFS and the second
derivative of GS energy shown in Eq.~(\ref{eq-fds-explicit}).

\subsection{Connection to quantum phase transitions}

In the dimerized model, as the two-spin RDMs with different parameters
commute, Eq.~(\ref{fs}) is applicable. By using the expression of the RDM
(see Eqs.~(\ref{eq-dimer-RDM-1})-(\ref{eq-elements-RDM})), after some
calculations, the RFS for the density matrix $\rho _{ij}$ is derived as%
\begin{eqnarray}
\chi _{ij} &=&\frac{4\left( \partial _{\alpha }\langle \mathbf{S}_{i}\cdot
\mathbf{S}_{j}\rangle \right) ^{2}}{\left( 3+4\langle \mathbf{S}_{i}\cdot
\mathbf{S}_{j}\rangle \right) \left( 1-4\langle \mathbf{S}_{i}\cdot \mathbf{S%
}_{j}\rangle \right) }  \notag \\
&=&\frac{3\left( \partial _{\alpha }\langle \sigma _{iz}\sigma _{jz}\rangle
\right) ^{2}}{4\left( 1+\langle \sigma _{iz}\sigma _{jz}\rangle \right)
\left( 1-3\langle \sigma _{iz}\sigma _{jz}\rangle \right) },  \label{eq-fds}
\end{eqnarray}%
which depends on both the spin correlator $\langle \sigma _{iz}\sigma
_{jz}\rangle $ itself and its first derivative. In fact, to ensure the
eigenvalues of $\varrho _{1}$ and $\varrho _{2}$ positive (we do not
consider the zero eigenvalues), it is required that
\begin{equation}
\langle \sigma _{iz}\sigma _{jz}\rangle \in (-1,\frac{1}{3}),
\label{eq-fds-condition}
\end{equation}%
which subsequently guarantees the susceptibility is non-negative.

Now, substituting Eq.~(\ref{spin-correlators}) into Eq.~(\ref{eq-fds}), one
can get another forms for the RFSs $\chi _{12}$ and $\chi _{23}$ as follows
\begin{eqnarray}
\chi _{12} &=&\frac{16\alpha ^{2}\left( \partial _{\alpha }^{2}e_{0}\right)
^{2}}{\left( 3+8e_{0}-8\alpha \partial _{\alpha }e_{0}\right) \left(
1-8e_{0}+8\alpha \partial _{\alpha }e_{0}\right) },  \notag \\
\chi _{23} &=&\frac{16\left( \partial _{\alpha }^{2}e_{0}\right) ^{2}}{%
\left( 3+8\partial _{\alpha }e_{0}\right) \left( 1-8\partial _{\alpha
}e_{0}\right) },  \label{eq-fds-explicit}
\end{eqnarray}%
in terms of GS energy and its first and second derivatives.

One key observation is that the numerators of the above two expressions
happen to be proportional to the square of the second derivative of GS
energy. Since the first derivative of energy is easily to be checked
continuous (see Eq.~(\ref{eq-Egs-derivative-infty})) and the denominators
are ensured to be positive and finite by Eq.~(\ref{eq-fds-condition}), the
singularities of the RFSs are determined only by the numerators. That is, if
the second derivative of GS energy is singular at the critical point, the
RFSs is singular too. On the other hand, it is known that the divergence of
the second derivative of GS energy reflects the second-order QPTs of the
system, which is shown in Ref. \cite{fidelity-energy-chen} explicitly as
\begin{equation}
\partial _{\alpha }^{2}e_{0}=\sum_{n\neq 0}^{N}\frac{2\left\vert \langle
\Psi _{n}|H_{1}|\Psi _{n}\rangle \right\vert ^{2}}{N(E_{0}-E_{n})},
\label{eq-2nd-derivative-Egs-QPT-chen}
\end{equation}%
where $H_{1}=\partial _{\alpha }H$ is the driving term of the Hamiltonian $H$%
, and $|\Psi _{n}\rangle $ is the eigenvector corresponding to the
eigenvalue $E_{n}$ of $H$. Eq.~(\ref{eq-2nd-derivative-Egs-QPT-chen}) shows
that the vanishing energy gap in the thermodynamic limit can lead to the
singularity of the the second derivative of GS energy. Therefore, both the
two-spin RFSs can exactly reflects the second-order QPTs in this model. In
addition, the second power in the numerators of the expressions, which
origins from the relation obtained in Eq.~(\ref{fs}), indicates that the
two-spin RFSs is more effective than the second derivative of the GS energy
in measuring QPTs. Furthermore, by the fidelity approach, it will be shown
in Sec. III that the dimerized AHC has a second-order critical point at $%
\alpha =1$.

\begin{figure*}[ptb]
\begin{center}
\includegraphics[height=6cm, width=7cm ] {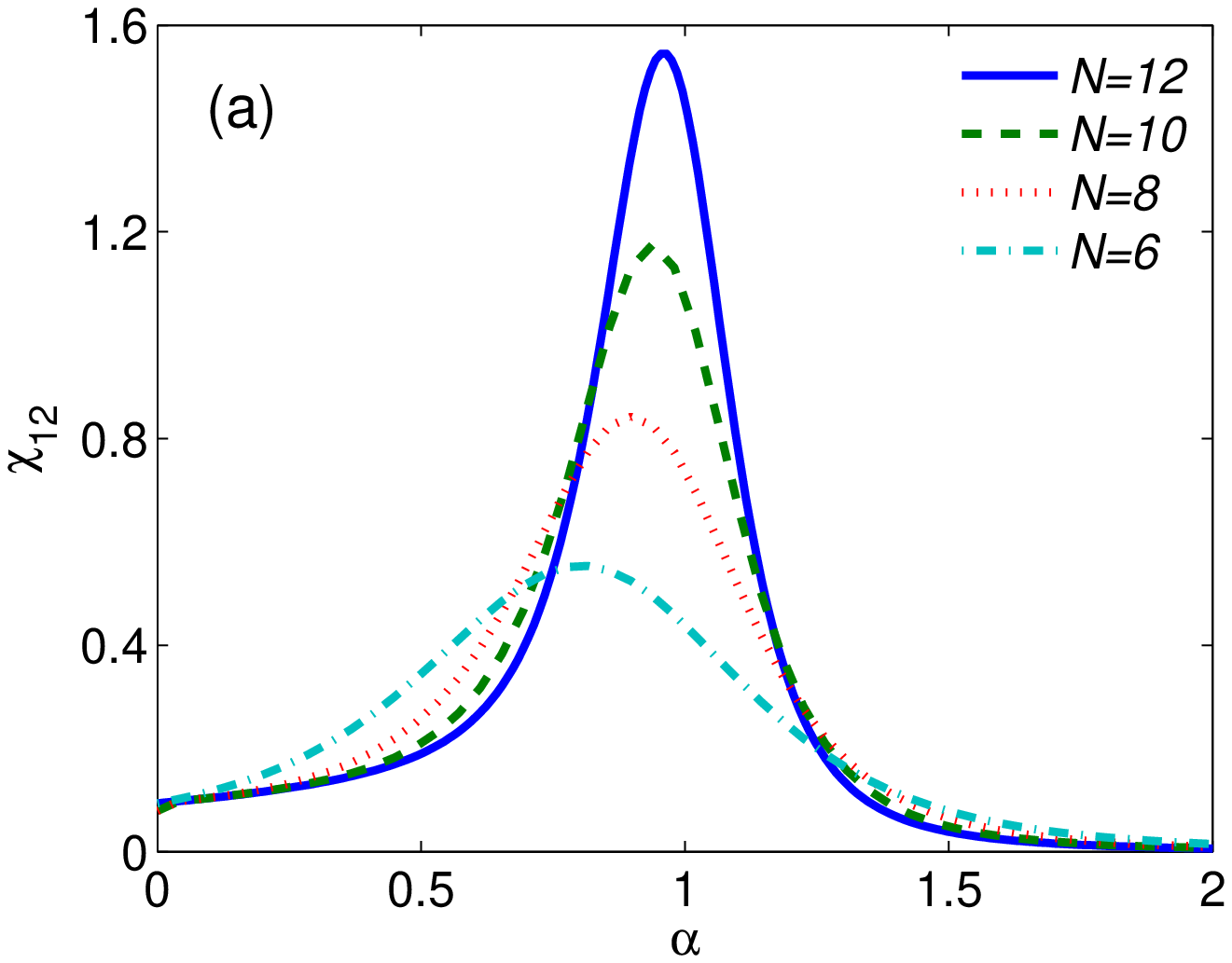} %
\includegraphics[height=6cm, width=7cm ] {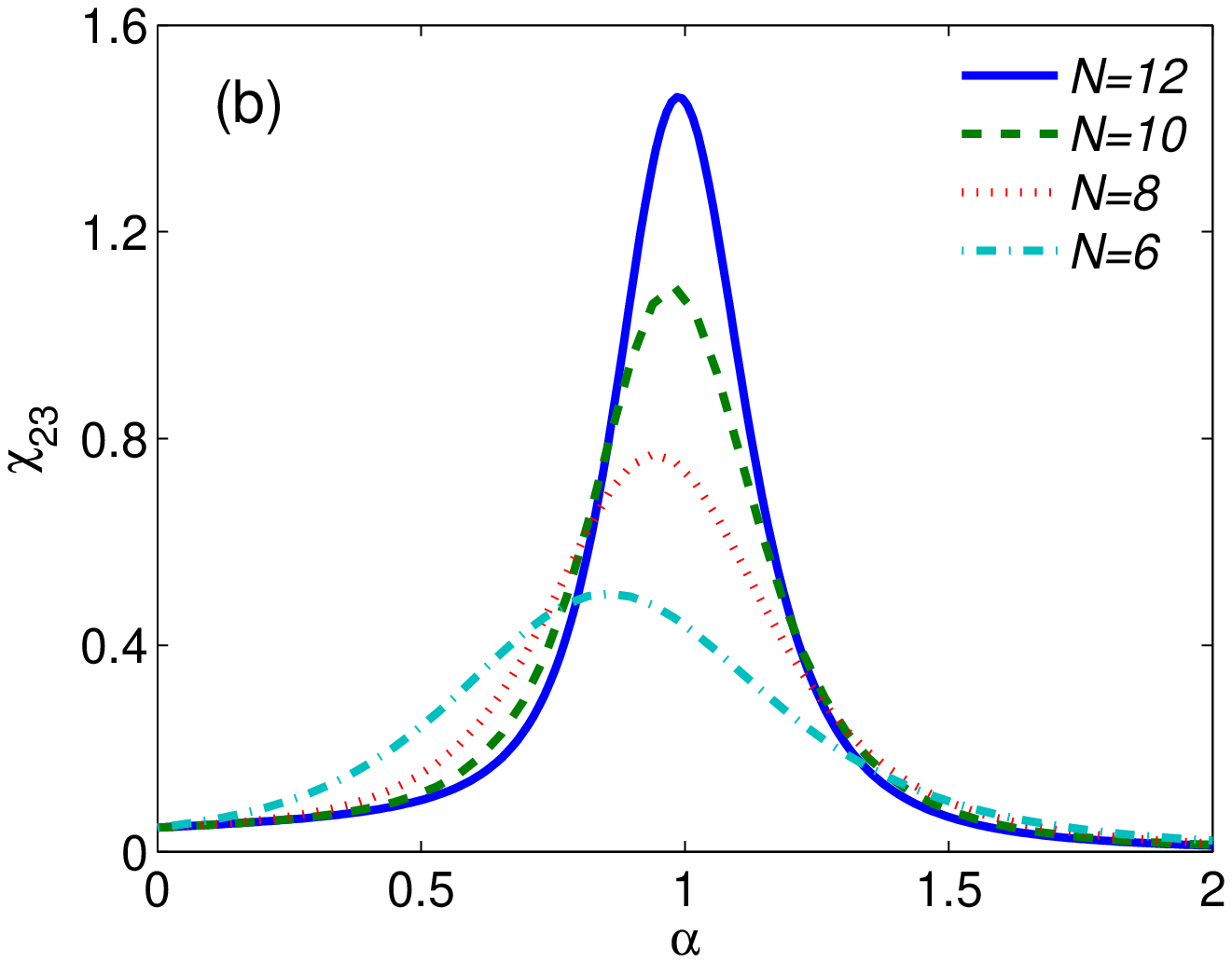}
\end{center}
\caption{Reduced fidelity susceptibilities (a) $\protect\chi_{12}$ and (b) $%
\protect\chi_{23}$ versus $\protect\alpha$ for $N=6,8,10,12$ for the
dimerized Heisenberg chain.}
\label{fig-fds123}
\end{figure*}

\section{Finite-size and critical behavior}

In this section, we consider the critical behavior of the 1D spin-1/2
dimerized AHC. It is known that, for $0<\alpha \ll {1}$, the coupling
between two dimers is so weak that all the spins are locked into singlet
states, while for $\alpha =1$, the system is reduced to the uniform AHC.
Hence, it has already been proved that the dimerized AHC has a critical
point at $\alpha =1$ \cite%
{DM-QPT-CUT,DM-QPT-DMRG,DM-QPT-concurrence,DM-QPT-entanglement}, which
exactly exists in the thermodynamic limit $N\rightarrow {\infty }$.

\subsection{Finite-size behavior}

\subsubsection{Analytical results for $N=4$ case}

For the case that the total spins $N=4$, the analytical results can be
obtained. In this case, the GS energy per spin of the system is \cite%
{DM-QPT-concurrence,GS energy}
\begin{equation}
e_{0}=-\frac{1}{4}\left( \frac{1+\alpha }{2}+\sqrt{1-\alpha +\alpha ^{2}}%
\right) ,  \label{eq-Egs}
\end{equation}%
with its first and second derivatives being
\begin{eqnarray}
\partial _{\alpha }e_{0} &=&\frac{1}{8}\left( -1+\frac{1-2\alpha }{\sqrt{%
1-\alpha +\alpha ^{2}}}\right) ,  \notag \\
\partial _{\alpha }^{2}e_{0} &=&-\frac{3}{16\left( 1-\alpha +\alpha
^{2}\right) ^{3/2}}.  \label{eq-Egs-4-derivative}
\end{eqnarray}%
Then the susceptibilities of the RDMs $\rho _{12}$ and $\rho _{23}$ can be
derived from Eq.~(\ref{eq-fds-explicit}) as
\begin{equation}
\chi _{12}=\chi _{23}=\frac{3}{16\left( 1-\alpha +\alpha ^{2}\right) ^{2}}.
\label{eq-fds-4}
\end{equation}

From Eq.~(\ref{eq-fds-4}) we see that $\chi _{12}$ and $\chi _{23}$ have the
same expressions, and there is no singularity over parameter $\alpha $.
However, take derivation of the expression with respect to $\alpha $, one
will find that there is a maximum of $\chi _{12}$ (or $\chi _{23}$) at $%
\alpha =0.5$, which is also the maximum position of $\partial _{\alpha
}^{2}e_{0}$ as shown in Eq.~(\ref{eq-Egs-4-derivative}). However, the
maximum position $\alpha =0.5$ deviates from the real critical point $\alpha
=1$ and can be called \textit{pseudo-critical point} due to the finite size
of the system. In addition, the different powers in the expressions of $\chi
_{12}$ (or $\chi _{23}$) and $\partial _{\alpha }^{2}e_{0}$ over the factor $%
(1-\alpha +\alpha ^{2})$, i.e., the former is $3/2$ and the latter is $2$,
shows that the RFS is more sensitive around the critical point.

Besides, the exact equivalence between $\chi _{12}$ and $\chi _{23}$ is in
contract with concurrences as shown in Ref. \cite{DM-QPT-concurrence}.
There, the concurrences for the reduced system, i.e., $C_{12}$ and $C_{23}$
are unequal to each other and have a crossing point at $\alpha =1$, which
leads to the mean concurrence takes its maximum at the critical point $%
\alpha =1$. This is because the concurrences $C_{12}$ and $C_{23}$ are only
related to the GS energy and its first derivative over $\alpha $,
respectively. However, the RFSs shown in Eq.~(\ref{eq-fds-explicit}) are
also determined by the second derivative of GS energy, which leads to the
identical behavior between $\chi _{12}$ and $\chi _{23}$.

\subsubsection{Numerical results for $N=6,8,10,12$}

For the case that the total spins $N>4$, we use exact diagonalization method
to examine the critical behavior of the system in terms of the RFSs. The
results for $N=6,8,10,12$ are shown in Fig.~\ref{fig-fds123}.

It is seen that both the RFSs $\chi_{12}$ and $\chi_{23}$ can well reflect
the critical behavior of the system. With increasing system size, the
pseudo-critical point exhibited by $\chi_{12}$ (or $\chi_{23}$) approaches
to the real critical point $\alpha=1$. Besides, the larger $N$ becomes, the
higher and shaper the peak of $\chi_{12}$ (or $\chi_{23}$) is.

It should be noticed that there is a slight difference between $\chi _{12}$
and $\chi _{23}$ for a given $\alpha $ and $N$, which results from the
difference between the spin correlators shown in Eq.~(\ref{spin-correlators}%
). In fact, the two spin correlators are equivalent, if we exchange the two
kinds of NN couplings. Thus $\chi _{12}$ and $\chi _{23}$ are also
equivalent in identifying QPTs.

\subsection{Infinite-size critical behavior}

Now, we consider the thermodynamic limit. To be consistent with the former
works, we adopt a new parameter $\eta \equiv {{(1-\alpha )}/{(1+\alpha )}}$.
When the system approaches to the uniform chain limit, i.e., $\eta
\rightarrow {0}$, analytical studies obtained by renormalization group \cite%
{EGS-infty-predict-1,EGS-infty-predict-2} had predicted that the GS energy
per spin ${e}_{0}$ should diverge as a power law times a logarithmic
correction, i.e., $\eta ^{4/3}/|\ln \eta |$. However, it is restricted to an
extremely small range $\eta <0.02$ \cite{DM-QPT-DMRG}. Thereafter, some
numerical results pointed out that a pure power-law behavior is reasonably
simple and accurate for larger $\eta $ as well \cite%
{EGS-infty-1,EGS-infty-2,DM-QPT-DMRG}.

\begin{figure}[ptb]
\begin{center}
\includegraphics[height=6cm, width=7cm ] {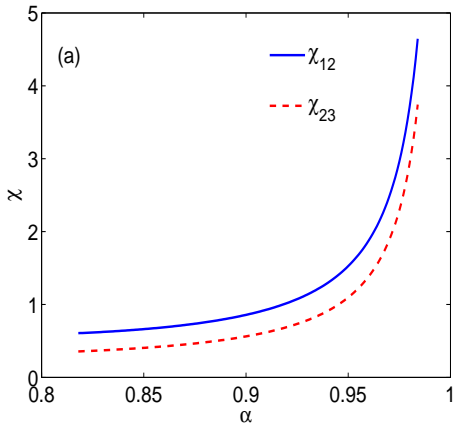}
\end{center}
\caption{Reduced fidelity susceptibilities versus $\protect\alpha$ in the
thermodynamic limit for the dimerized Heisenberg chain, with $c=0.3891$ and
the exponent $p=1.4417$ over the range of ${0.001}\leq\protect\eta\leq{0.1}$
\protect\cite{EGS-infty-2}, i.e., ${0.8182}\leq\protect\alpha\leq{0.9880}$.}
\label{fig-fds-limit}
\end{figure}

For generality, we assume a power law of ${e}_{0}$ as the form $c{\eta ^{p}}$
with $c$ an overall constant. The exponent $p$ are given differently over
different $\eta $ ranges. Hitherto, almost all the works \cite%
{EGS-infty-1,EGS-infty-2,DM-QPT-DMRG} show that $1<p<2$ over the range $%
0<\eta <1$. For example, using the DMRG approach, in \cite{DM-QPT-DMRG}, the
exponent is fit to be $p=1.45$ over the range of ${0.008}\lesssim \eta
\lesssim {0.1}$ with $c=0.39$, and in Ref.~\cite{EGS-infty-2}, it is
estimated in the range of ${{0.001}\leq }\eta {\leq {0.1}}$ as that $p=1.4417
$ with $c=0.3891$. Thus we will restrict $1<p<2$ in the following. The GS
energy per spin in the thermodynamic limit can be written accordingly as
\cite{DM-QPT-DMRG}
\begin{equation}
{e}_{0}(\eta )=\frac{1}{1+\eta }(e_{0}(0)-c{\eta ^{p}}),
\label{eq-Egs-infty}
\end{equation}%
where $e_{0}(0)=1/4-\ln {2}$ is the GS energy per spin for $\eta =0$.

The above expression shows that the GS energy follows the power law behavior
$\eta ^{p}$. This gives a prediction of the critical point of the RFSs. From
Eq.~ (\ref{eq-Egs-infty}), we can easily get the first and second
derivatives of GS energy per spin in the thermodynamic limit as
\begin{eqnarray}
\partial _{\alpha }{e}_{0} &=&\frac{c}{2}\left( 2p+\alpha -1\right) \left(
1+\alpha \right) ^{-1}\left( 1-\alpha \right) ^{p-1},  \notag
\label{eq-Egs-derivative} \\
\partial _{\alpha }^{2}{e}_{0} &=&-2c\left( p-1\right) p\left( 1+\alpha
\right) ^{-\left( p+1\right) }\left( 1-\alpha \right) ^{p-2}.
\label{eq-Egs-derivative-infty}
\end{eqnarray}%
It is seen that, as ${\alpha }>0$ and $1<p<2$, the first derivative of ${e}%
_{0}$ does not diverge for any allowed $\alpha $ value, while the second
derivative of ${e}_{0}$ has a singular point $\alpha =1$. According to Eq.~(%
\ref{eq-fds-explicit}), it is no doubt that the RFSs also diverges at $%
\alpha =1$. That is, the dimerized AHC has a second-order critical point $%
\alpha =1$.

Next we discuss the critical behavior of the RFSs around the critical point.
Insert Eq.~(\ref{eq-Egs-derivative}) into Eq.~(\ref{eq-fds-explicit}), we
obtain the RFSs as
\begin{widetext}
\begin{eqnarray}
\chi _{12} &=&-\frac{c^{2}p^{2}\left( p-1\right) ^{2}\eta
^{-2+2p}\left( \eta -1\right) ^{2}\left( \eta +1\right)
^{4}}{16\left[ c^{2}\left( p+\eta -p\eta \right) ^{2}\eta
^{2p}+c\left( 2\ln 2-1\right) \left( p+\eta -p\eta \right) \eta
^{1+p}+\ln 2\left( \ln 2-1\right) \eta ^{2}\right] }, \nonumber\\
 \chi _{23}
&=&-\frac{c^{2}p^{2}\left( p-1\right) ^{2}\eta ^{-2+2p}\left( \eta
+1\right) ^{6}}{16\left[ c^{2}\left( p-\eta +p\eta \right) ^{2}\eta
^{2p}-c\left( 2\ln 2-1\right) \left( p-\eta +p\eta \right) \eta
^{1+p}+\ln 2\left( \ln 2-1\right) \eta ^{2}\right] }.
\label{eq-RFSs-infty}
\end{eqnarray}
\end{widetext}
When $\alpha\rightarrow{1}$, i.e., $\eta\rightarrow{0}$, we only consider
the leading terms in the expressions and get the critical behavior of the
RFSs as
\begin{equation}
\chi _{12},\chi _{23}\sim {\eta^{2p-4}}\sim(1-\alpha)^{2p-4}.
\label{eq-RFSs-infty-critical}
\end{equation}
Obviously, for $1<p<2$, both of them diverge at $\eta=0$, i.e., $\alpha=1$,
as displayed in Fig. \ref{fig-fds-limit}. It is shown that the two RFSs
diverge quickly when $\alpha$ approaches to $1$. For a given $\alpha$, $%
\chi_{12}$ and $\chi_{23}$ are remarkably larger than those in the
finite-size cases. In addition, the different power between $\partial
_{\alpha }^{2}e_{0}$ and $\chi _{12} (\chi _{23})$ over the factor $%
(1-\alpha)$ indicates that these RFSs are more singular around the critical
point.

\section{General connection between reduced fidelity and quantum phase
transitions}

\subsection{More exemplifications}

In the above, we have illustrated the connection between RFS and QPTs in the
spin-$1/2$ dimerized AHC, which has SU(2) symmetry. Actually, it is
straightforward that the RFS expression (\ref{eq-fds}) is general for an
arbitrary spin-$1/2$ Hamiltonian with SU(2) symmetry. Combined with the
translation symmetry, it is easy to obtain a relation between RFS and QPTs
like Eq.~(\ref{eq-fds-explicit}). In the following, we would like to give
another two exemplifications, which are also of SU(2) and translation
symmetries. One is the mixed-spin ($1/2$, $S$) dimerized Heisenberg chain
with $S$ an arbitrary spin length, the other one is the spin-1
bilinear-biquadratic model.

The Hamiltonian for the mixed-spin dimerized Heisenberg chain with
alternated spins $\mathbf{S}_{1}$ and $\mathbf{S}_{2}$ is
\begin{equation}
H_{F}=\sum_{i=1}^{N/2}\left( \mathbf{S}_{1,i}{\cdot }\mathbf{S}_{2,i}+\alpha
\mathbf{S}_{2,i}{\cdot }\mathbf{S}_{1,i+1}\right) ,
\label{eq-Hamiltoonian-mixed-dimer}
\end{equation}%
here, $\mathbf{S}_{1}$ and $\mathbf{S}_{2}$ denote the spin-$1/2$ and spin-$%
S $ operators respectively, and $\alpha $ is the ratio between the two kinds
of NN spin couplings. The periodic boundary condition is assumed. As the
system is of SU(2) symmetry, the RDM between two NN coupling spins can be
expressed in the coupled angular momentum representation as \cite%
{arbitrary-SU(2)-invariant-density-matrix}

\bigskip
\begin{eqnarray}
\rho _{ij} &=&\frac{F}{2S}\sum_{J_{z}=-S+1/2}^{S-1/2}|S-1/2,J_{z}\rangle
\langle S-1/2,J_{z}|  \notag  \label{eq-RDM-mixed-dimer} \\
&&+\frac{1-F}{2S+2}\sum_{J_{z}=-S-1/2}^{S+1/2}|S+1/2,J_{z}\rangle \langle
S+1/2,J_{z}|,  \notag \\
&&
\end{eqnarray}%
with $J_{z}$ the total spin along the $z$ direction of the two spins and $%
F=\left( S-2\langle \mathbf{S}_{1,i}{\cdot }\mathbf{S}_{2,j}\rangle \right)
/(2S+1)$. Since $\rho _{ij}$ is diagonal, the RFS expression (\ref{fs}) is
available. Then
\begin{equation}
\chi _{ij}=\frac{\left( \partial _{\alpha }\langle \mathbf{S}_{1,i}{\cdot }%
\mathbf{S}_{2,j}\rangle \right) ^{2}}{\left( S-2\langle \mathbf{S}_{1,i}{%
\cdot }\mathbf{S}_{2,j}\rangle \right) \left( S+1+2\langle \mathbf{S}_{1,i}{%
\cdot }\mathbf{S}_{2,j}\rangle \right) }.  \label{eq-RFS-mixed-dimer-general}
\end{equation}%
Meanwhile, the system is translational invariant. Thus applying the
Feynman-Hellman theorem to the GS of the system, we get the expressions for
the two kinds of RFSs between two NN spin pairs
\begin{eqnarray}
\chi _{12} &=&\frac{4\alpha ^{2}\left( \partial _{\alpha }^{2}e_{0}\right)
^{2}}{\left( S-4e_{0}+4\alpha \partial _{\alpha }e_{0}\right) \left(
S+1+4e_{0}-4\alpha \partial _{\alpha }e_{0}\right) },  \notag \\
\chi _{23} &=&\frac{4\left( \partial _{\alpha }^{2}e_{0}\right) ^{2}}{\left(
S-4\partial _{\alpha }e_{0}\right) \left( S+1+4\partial _{\alpha
}e_{0}\right) }.  \label{eq-RFS-mixed-dimer-explicit}
\end{eqnarray}%
Obviously, when $S=1/2$, the above expression reduces to Eq. (\ref%
{eq-fds-explicit}). The two RFSs is proportional to the second derivative of
the GS energy per spin $e_{0}\equiv {e_{0}(\alpha )}$. That is, the RFS also
has the possibility to signal the second-order QPTs of a mixed-spin system.

Furthermore, the general expression of RFS~(\ref{eq-fds}) could also be
extended to high-spin system, like the spin-$1$ bilinear-biquadratic model,
which describes the structure of some materials, such as LiVGe$_{2}$O$_{6}$%
\cite{LiVGe2O6-1,LiVGe2O6-2}. The Hamiltonian reads
\begin{equation}
H_{BB}=\sum_{i=1}^{N}\left[ \cos {\theta }(\mathbf{S}_{i}{\cdot }\mathbf{S}%
_{i+1})+\sin {\theta }(\mathbf{S}_{i}{\cdot }\mathbf{S}_{i+1})^{2}\right] ,
\label{eq-Hamiltoonian-BB}
\end{equation}%
here $\mathbf{S}_{i}$ denotes the spin-1 operator, and $\theta $ reflects
the different coupling strengths. The periodic boundary condition is assumed
as well. Obviously, this Hamiltonian is also of SU(2) and translation
symmetries. In Eq.~(24) of Ref.~\cite{RFS-higher-spins}, the QPT of this
model is studied by using the RFS between NN-coupling spins, which happens
to be proportional to the second derivative of the GS energy density $e_{0}{%
\equiv }e_{0}(\theta )$, i.e.,
\begin{equation}
\chi _{12}\propto {(e_{0}+\partial _{\theta }^{2}{e_{0}})^{2}}.
\label{eq-RFS-bb-model-explicit}
\end{equation}%
This further confirms that the two-spin RFS is an effective tool to reveal
the second-order QPTs even for high-spin systems.

All the above results show that the two-spin RFS is tied to the
corresponding spin-correlator, while the latter is related to the GS energy
through Feynman-Hellman theorem. Then the RFS is connected to the square of
the second derivative of the GS energy, which is used to characterize the
second-order QPTs, as shown in Eqs.~(\ref{eq-fds-explicit}),~(\ref%
{eq-RFS-mixed-dimer-explicit}),~(\ref{eq-RFS-bb-model-explicit}). Moreover,
the square relation between the two-body RFS and the second derivative of GS
energy Eq.~(\ref{eq-fds-explicit}) holds for a broad class of systems with
SU(2) and translaton symmetries, and thus the RFS might be more sensitive
than... the second derivative of the GS energy in characterizing QPTs.

\subsection{General connection between reduced fidelity and quantum phase
transitions}

The models considered above are all of SU(2) and translation symmetries. It
is noticed that the definition of RF (\ref{eq-definition-mixed-fidelity})
depends only on the RDM, which may contains sufficient information about
QPTs. This inspires us to infer that for a more general case, QPTs are
essentially related to the RDM. In \cite{RMD-QPT}, they have provided a
powerful substantiation. They demonstrated that, under certain general
conditions, the elements of two-body RDM are able to signal the QPTs. They
consider a general Hamiltonian that contains two-body interaction like
\begin{equation}
H=\sum_{i{\alpha }{\beta }}{\epsilon }_{\alpha \beta }^{i}{|\alpha
_{i}\rangle }{\langle \beta _{i}|}+\sum_{ij{\alpha }{\beta }\gamma \kappa
}V_{\alpha \beta \gamma \kappa }^{ij}{|\alpha _{i}\rangle }{|\beta
_{j}\rangle }{\langle \gamma _{i}|}{\langle \kappa _{j}|},
\end{equation}%
where $i$, $j$ enumerate $N$ particles and $\{|\alpha _{i}\rangle \}$ is a
basis for the Hilbert space. For the nondegenerate GS $|{\psi }\rangle $,
its GS energy is $E_{0}=\langle {\psi }|H|{\psi }\rangle $, and the element
of the corresponding two-particle RDM is $\rho _{\gamma \delta \alpha \beta
}^{ij}=\langle {\psi }|\alpha _{i}\beta _{j}\rangle \langle \gamma
_{i}\kappa _{j}|\psi \rangle $. Thus the relation between energy and RDM is $%
E_{0}=\sum_{ij}$tr$[U(ij)\rho ^{ij}]$, where $U_{\alpha \beta \gamma \kappa
}(ij)=\epsilon _{\alpha \gamma }^{i}\delta _{\beta \kappa
}^{j}/N_{i}+V_{\alpha \beta \gamma \kappa }^{ij}$ with $N_{i}$ the number of
particles that particle $i$ interacts with and $\delta _{\beta \kappa }^{j}$
the Kronecker symbols on particle $j$. Then using the Feynman-Hellman
theorem, the derivatives of energy per particle $(e_{0}{\equiv }E_{0}/N)$
are obtained as
\begin{eqnarray}
\partial _{\xi }e_{0} &=&\frac{1}{N}\sum_{ij}\text{tr}[(\partial _{\xi
}U(ij))\rho _{ij}],  \label{eq-RDM-1QPT} \\
\partial _{\xi }^{2}e_{0} &=&\frac{1}{N}\sum_{ij}\{\text{tr}[(\partial _{\xi
}^{2}U(ij))\rho _{ij}]  \notag \\
&&+\text{tr}[(\partial _{\xi }U(ij))\partial _{\xi }\rho _{ij}]\},
\label{eq-RDM-2QPT}
\end{eqnarray}%
where it follows from Eq.~(\ref{eq-RDM-1QPT}) that $\sum_{ij}$tr$%
[U(ij)\left( \partial _{\xi }\rho _{ij}\right) ]$$=0$. As is known,
according to the classical definition of phase transitions given in terms of
the free energy \cite{book-free-energy-QPT}, in the limit of $T=0$, a
first-order QPT (second-order QPT) is characterized by a discontinuity in
the first (second) derivative of the GS energy (see also Eq.~(\ref%
{eq-2nd-derivative-Egs-QPT-chen})). Therefore, if $U(ij)$ is a smooth
function of the Hamiltonian parameter $\xi $, the origin of first-order QPTs
is due to the discontinuity of one or more of the $\rho _{ij}$'s at the
critical point according to Eq.~(\ref{eq-RDM-1QPT}). Whereas, if $\rho _{ij}$
is finite at the critical point, the origin of second-order QPTs is the fact
one or more of the $\partial _{\alpha }\rho _{ij}$'s diverge at the critical
point.

Based on these facts, one find that if $U(ij)$ is a smooth function and the
first\ derivative of the elements of $\rho _{ij}$ diverges at the critical
point, then $\partial _{\xi }^{2}{e_{0}}$ diverges too, which indicates a
second-order QPT. For example, in our models considered, the elements of $%
\rho _{ij}$ are decided by $\langle \sigma _{iz}\sigma _{jz}\rangle $, which
is connected to $\partial _{\xi }{e_{0}}$ via Feynman-Hellman theorem. Thus
the relation revealed by Eq.~(\ref{eq-RDM-2QPT}) may be the origin of the
relation between RFS and QPTs. This is not restricted to the systems with
SU(2) invariance, and a more explicit and direct relation between RFS and
QPTs may need further deep considerations.

In addition, the relation between the reduced fidelity (denoted as $F_{R}$)
and its corresponding global fidelity $F_{G}$ is given already as $F_{G}\leq
{F_{R}}$ \cite{relation-global-reduced-fidelity}. According to relation (\ref%
{eq-definition-fidelity-susceptibility}), the corresponding susceptibilities
satisfy $\chi _{G}\geq \chi _{R}$. However, all the previous works \cite%
{averaged-fidelity-zhou,averaged-fidelity-zhou-XYmodel,partial-fidelity,partial-fidelity-LMG-model,reduced-fidelity-LMG-Ma,reduced-fidelity-TIsing-Ma}
and this work confirm that the reduced fidelity approach is as effective as
global fidelity in characterizing QPTs, and in some cases, such as the
models mentioned above, it is only necessary to know the GS energy of system
in calculating the RFS, rather than its GS for the global fidelity, which is
generally not easy to be obtained.

\section{Conclusion}

In conclusion, we have studied the critical behavior of the 1D spin-1/2
antiferromagnetic Heisenberg chain with dimerization in terms of RFS. For
the GS of the system, two kinds of RFSs between two NN spin pairs are
considered. It is interesting that, due to the SU(2) and translation
symmetries, the singularities of these RFSs are just determined by the
square of the second derivative of the GS energy, which means the RFS is
more effective than the second derivative of the GS energy in identifying
the second-order QPTs. Explicit calculations are also carried out both in
finite-size and infinite-size situations. It is found that, as the system
size increases, the pseudo-critical points of the RFSs approach to the real
critical point $\alpha =1$. In the thermodynamic limit, we give the critical
exponent of the two RFSs. These results further convince us that the
critical behavior of the system can be reflected by the fidelity of its
two-spin subsystem, which is of practical use in experiments.

Furthermore, concluding the results obtained in the spin-1/2 dimerized
model, we examine another two examples, i.e., the mixed-spin dimerized
Heisenberg chain and the spin-1 bilinear-biquadratic model, which have the
SU(2) and translation symmetries as well. It is also found that the RFSs are
directly connected to the square of the second derivative of the GS energy,
which indicates that for a broad class of systems with SU(2) and translation
symmetries, the RFS is more effective than the second derivative of the GS
energy in reflecting second-order QPTs. Moreover, the origin between RFS and
QPTs is found to be generally rooted in the relations between the elements
of the RDM and derivatives of GS energy, which provides a general proof for
the fact that RFS can be used to detect the second-order QPTs of the system.

\textit{Acknowledgements} This work was supported by the Program for New
Century Excellent Talents in University (NCET), the NSFC with grant No.
90503003, the State Key Program for Basic Research of China with grant No.
2006CB921206, the Specialized Research Fund for the Doctoral Program of
Higher Education with grant No. 20050335087.

\end{document}